\documentclass[lettersize,journal]{IEEEtran}
\usepackage{amsmath,amsfonts}
\usepackage{algorithmic}
\usepackage{algorithm}
\usepackage{array}
\usepackage[caption=false,font=normalsize,labelfont=sf,textfont=sf]{subfig}
\usepackage{soul}
\usepackage{braket}
\usepackage{textcomp}
\usepackage{stfloats}
\usepackage{physics}
\usepackage{url}
\usepackage{verbatim}
\usepackage{graphicx}
\usepackage{cite}
\usepackage{siunitx}
\usepackage{subcaption}
\usepackage{graphicx} 
\begin{document}

\title{Backflash attack on Coherent One-Way Quantum Key Distribution}

\author{Ashutosh Kumar Singh, Nilesh Sharma, Vaibhav Pratap Singh, Anil Prabhakar         
\thanks{AKS, NS, VPS, AP are with the Centre for Quantum Information, Communication and Computing~(CQuICC), Indian Institute of Technology Madras, and also with the Department of Electrical Engineering, Indian Institute of Technology Madras, India. (email: ashutosh@ee.iitm.ac.in, nileshlns@smail.iitm.ac.in, ee20d036@smail.iitm.ac.in, anilpr@ee.iitm.ac.in)}
\thanks{VPS is also affiliated with the Centre for Development of Advanced Computing Bangalore~(CDAC-B), India.}

}

\markboth{Placeholder}
{Shell \MakeLowercase{\textit{et al.}}: A Sample Article Using IEEEtran.cls for IEEE Journals}

\maketitle

\begin{abstract}
In this article, we experimentally demonstrate an eavesdropper’s (Eve’s) information gain by exploiting the breakdown flash generated by the single photon avalanche detector (SPAD) used in coherent one-way quantum key distribution (COW-QKD) setup. Unlike prior studies focusing on the device-level characterization of backflash photons, this work quantifies Eve’s learning with a QKD system that includes a key distillation engine (KDE). Eve's learning is quantified using the ``Backflash'' photons emitted by SPAD and the information available on the classical channel. Experimentally observed data are in good agreement with the theoretical simulations. Some mitigation strategies against the backflash attack are also discussed.

\end{abstract}


\section{Introduction}
\IEEEPARstart{T}{he} prevalent class of cryptography relies on the hardness of certain computational problems to generate secure keys between multiple parties. The discovery of Shor's algorithm in 1994 highlighted the infirmity of this approach~\cite{shor94}. Various methods have been suggested to tackle this problem and make the encryption \textit{quantum proof} \cite{10392193}. Quantum key distribution (QKD) is one of them. In the first QKD proposal, two parties, colloquially referred to as Alice and Bob, are connected via two channels - a quantum channel for sharing quantum states and an authenticated classical channel for classical communication~\cite{BB84}. There have been subsequent developments in terms of new protocols to ease the experimental overhead whilst closing different vulnerabilities \cite{Stucki:09, Inoue02, Ranu_2021, sharma2024mitigating, sharma2024twin}.

The argument of information-theoretic security guaranteed by QKD holds true only if the practical devices behave according to their theoretical models. Deviation from such models leads to different types of attacks, either based on information leakage at the detector (side-channel attacks) or improper generation/handling of encoded qubits \cite{brassard2000limitations, makarov2005faked, makarov2009controlling}.

In this article, we experimentally demonstrate Eve's learning by employing the backflash attack on Bob's Single Photon Avalance Detector (SPAD) in a COW-QKD setup. The phenomena of emission of backflash photons was first reported by Newman in 1955 for Si p-n junction~\mbox{\cite{Newman}}. In the near-band-energy region, the emission has been attributed to the radiative recombination of an electron and a free hole generated during the self-sustaining avalanche triggered due to the operation of SPAD in Gieger mode~\mbox{\cite{Manfredi_Si}}. Prior work on studying the backflash have focused on exploring the temporal and spectral content of Silicon and InGaAs/InP SPADs~\cite{Acerbi_InGaAs, Manfredi_Si,Shi:17}.  
Both free-space and fibre-pigtailed SPADs have been used to emulate a QKD receiver to demonstrate Eve's learning from backflash attack using various techniques such as an optical time domain reflectometer (OTDR) working at single photon levels, or coincidence \mbox{electronics~\cite{Meda2017, Pinheiro:18, Shi:17, Vybornyi2021}}. Backflash attack in fast-gated avalanche photodiode (APD) has also been studied, and the information leakage has been compared against slow-gated APD for \mbox{QKD~\cite{Koehler-Sidki2020}}. These offer valuable insight into the phenomenon of backflash; however, they are quantifying Eve's learning on the basis of device characterisation. To the best of our knowledge, our work demonstrates the first attempt at using the backflash photons in an experimental COW-QKD setup that includes a key distillation engine~(KDE). We utilise the information available on the classical authenticated channel to experimentally demonstrate Eve's learning and compare it against theoretical estimations.

The article is organized into four sections. Section~2 details the experimental setup, while Section~3 presents the experimental results, their comparison with the estimated theoretical results and discussions. Finally, Section~4 provides some steps to mitigate against the attack along with the concluding remarks. 

\section{Experimental Setup}\label{section:experimental-setup}
\begin{figure}[h]
    \centering
    \includegraphics[width=\linewidth]{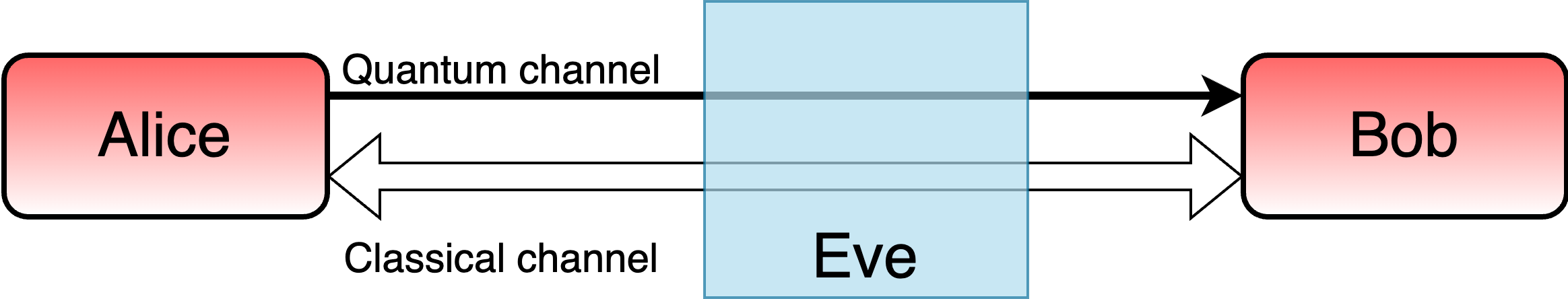}
    \caption{Block diagram of a two-party QKD system in the presence of an Eavesdropper.}
    \label{fig: block}
\end{figure}

A block-level diagram of a QKD system is shown in Fig.~\ref{fig: block}, where two legitimate parties, Alice and Bob, intend to extract symmetric, secure keys in the presence of Eve.  
In the security analysis of QKD protocols, the eavesdropper is typically assumed to possess omnipotent technological capabilities \cite{shor2000simple, branciard2008upper}, making her significantly more advanced than the legitimate parties. To account for Eve's assumed technological superiority, our experiment employs a Superconducting Nanowire Single-Photon Detector (SNSPD) from Single Quantum. Meanwhile, for key distribution between the legitimate parties, we utilise a commercially available InGaAs SPAD (PDM-IR) from Micro Photon Devices.

\begin{figure}[ht]
    \centering
    \includegraphics[width=0.45\linewidth]{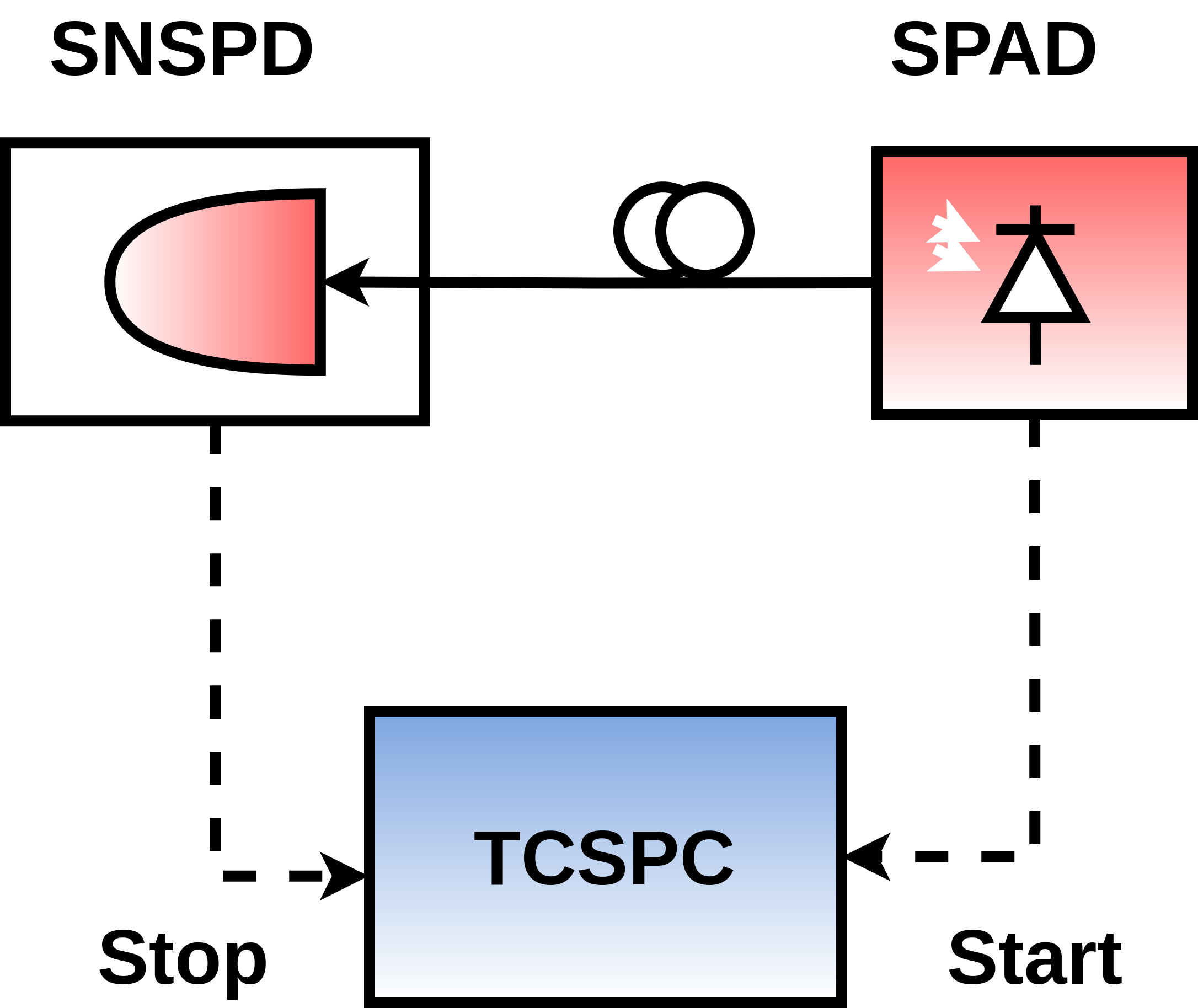}
    \caption{Experimental setup to observe timing correlation between a detection on SPAD and emitted backflash photons. TCSPC: Time-Correlated Single Photon Counting Module. Dashed lines represent RF connections.}
    \label{fig:backflash-timing-correlation}
\end{figure}

The timing correlation of the backflash photon had to be verified before the Eavesdropper could exploit it in her attack. The setup, as shown in Fig.~\ref{fig:backflash-timing-correlation}, shows an SPAD connected to an SNSPD via a single mode optical patch-cord. As backflash photons are generated due to the recombination of electron-hole pairs during the avalanche, timing correlation measurement can be performed without an input optical pulse~\cite{Manfredi_Si,Meda2017}. The SPAD is highly user-configurable in terms in terms of gate width, excess bias voltage, hold-off time, etc. For timing correlation, the hold-off time was set to 1~$\mu\text{s}$, with an excess bias voltage of 5~V, and the gate width was varied. The electrical output of both detectors is connected to a time tagging module (TCSPC) from qutools that records the arrival time of photons.

\begin{figure*}
    \centering
    \includegraphics[width=0.65\linewidth]{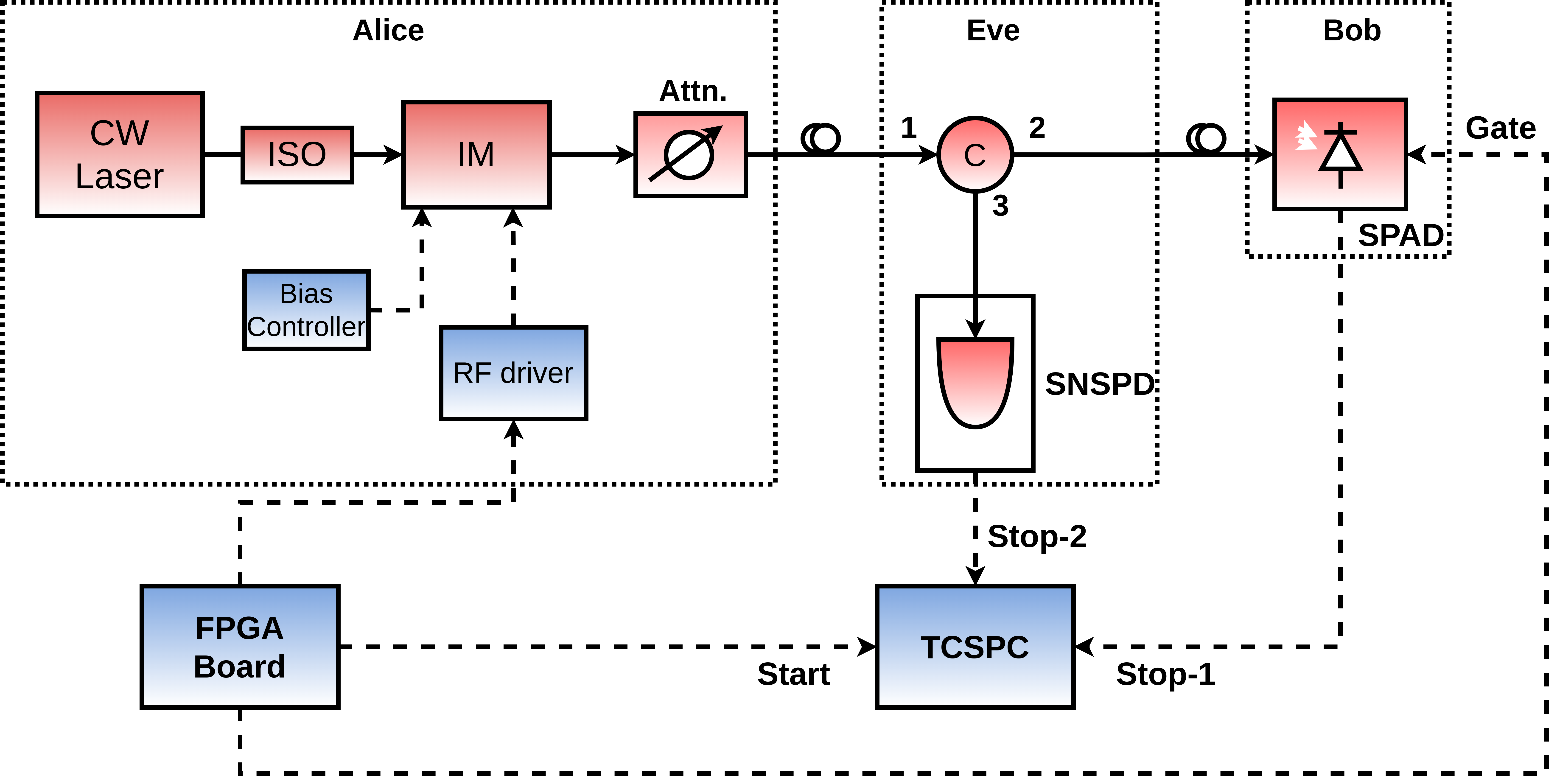}
    \caption{Experimental setup integrating Alice, Bob, and Eve. CW Laser: Continous-Wave Laser, ISO: Optical Isolator, IM: Intensity Modulator, Attn.: Optical Attenuators, C: Circulator, TCSPC: Time-Correlated Single Photon Counting Module. Solid lines indicate optical connections, and dashed lines indicate electrical/RF connections.}
    \label{fig:eve-setup}
\end{figure*}

We have implemented COW-QKD protocol as per the schematic shown in Fig.~\ref{fig:eve-setup}. COW belongs to a class of distributed phase reference QKD protocols where the key is derived based on the arrival time of photons~\cite{Stucki:09}. Alice uses a weak coherent source to prepare her two logic bits as $0_{\text{L}}=\ket{\alpha}_1\otimes\ket{0}_2$ and $1_\text{L}=\ket{0}_1\otimes\ket{\alpha}_2$, where $\ket{\alpha}$ represents a coherent pulse with mean photon number $\lvert\alpha\rvert^2$ and $\ket{0}$ represents a vacuum pulse. The subscript in the quantum state represents the time-bin. The security of COW-QKD is ensured by sending a decoy sequence, $\text{D}=\ket{\alpha}_1\otimes\ket{\alpha}_2$ from Alice and measuring the output of the destructive port of an unbalanced Mach–Zehnder interferometer at Bob. A conventional COW-QKD setup will employ a beamsplitter that multiplexes the data and monitoring line. Any backflash photons from the SPAD will retrace the same path along the data line and will be reduced by the insertional loss of the beamsplitter. Furthermore, since backflash attacks are passive in nature and do not disturb the coherence of the pulse, they differ fundamentally from more active threats such as photon number splitting attacks or faked state \mbox{attacks~\cite{Lütkenhaus_pns_2002,vadim_fakes_states}}. This makes the setup shown in \mbox{Fig.~\ref{fig:eve-setup}} well-suited for the scope of our analysis.

Experimentally, Alice realises her setup using a 1550~nm continuous-wave laser modulated by an external Lithium Niobate intensity modulator~(IM). The modulator is biased at the null point and driven by RF pulses from FPGA. A non-inverting pulsed RF amplifier is used to ensure the electrical pulse output voltage is closer to V$_\pi$ of the IM. In our implementation, the RF signal from FPGA is generated at a nominal rate of 1~GHz, translating into a time-bin width of 1~ns. The optical pulses are further attenuated to achieve a mean photon number, {$|\alpha|^2 < 1$}. Eve's setup consists of an SNSPD and an optical circulator. SNSPD ensures low DCR and higher photon detection efficiency, giving Eve a technological advantage over Bob. To demonstrate backflash attack, Alice sends two logical bits in a 4~ns window consisting of $0_\text{L}$ in the first half, and $1_\text{L}$ in the second half, padded by vacuum states repeating every 32~ns.
The limit of low duty cycle is dictated by the pulsed RF amplifier used in our setup. We have developed Bob with only one SPAD for the detection line. For the experimental setup characterising Eve's learning, the SPAD is operated in gated mode triggered by an RF signal from the FPGA. The gate width (T$_\text{ON}$) is set as 4~ns, repeating every 32~ns. The hold-off time is set as 10~$\mu\text{s}$ to reduce the dark count rate~(DCR) and after-pulsing effect.

In a field-deployed QKD scheme, Alice and Bob would be situated at different geographical locations with a quantum channel, an authenticated classical channel, and a clock-synchronisation setup. We assume that Eve's omnipotence allows it to place the circulator outside Bob's secure boundary and synchronise its clock. With these assumptions, we can reduce the system complexity by using only one FPGA board and a TCSPC module.
\section{Results and Discussion}
In the COW-QKD protocol, the information is encoded in the time-bins and decoded at Bob, depending on the arrival time of photons. Consequently, Eve's ability to extract logical bits will depend on the timing correlation between the SPAD clicks at Bob and the subsequent release of backflash photons detected at Eve's SNSPD. 
\subsection{Timing Correlation}\label{Sub: Timing Correlation}
\begin{figure}[h]
    \centering
    \includegraphics[width=0.8\linewidth]{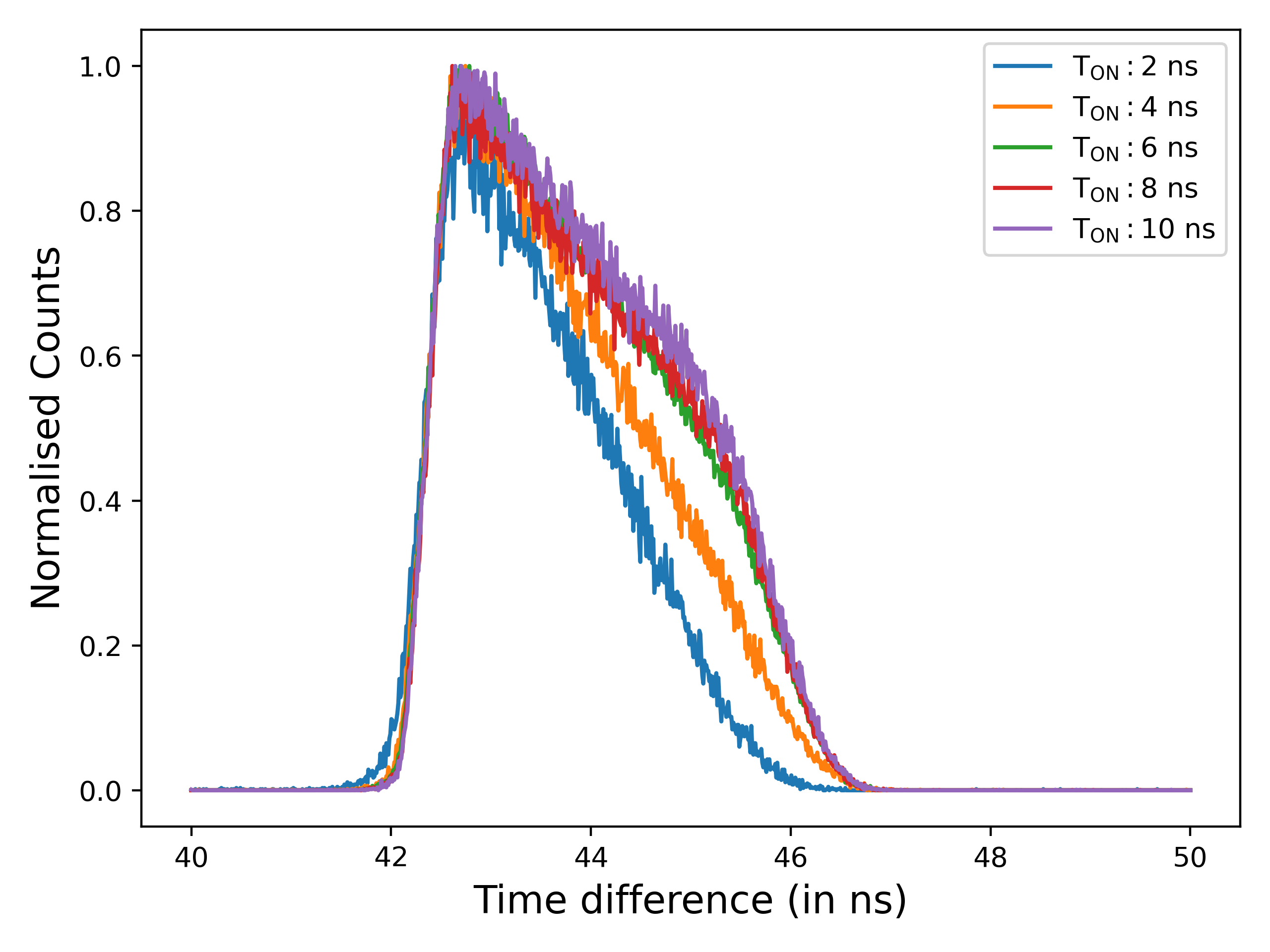}
    \caption{Histogram generated with bin-width 10~ps for time difference between SPAD and SNSPD detections for different SPAD gate widths.}
    \label{fig:timing-correlation}
\end{figure}
The state of the photons emitted due to backflash is not expected to carry any information about the state of incident photon~\cite{Pinheiro:18}. However, it is important to verify its temporal correlation. We measured the DCR of SNSPD to be around $16.4\pm3.5$~cps when operated at the photon detection efficiency of about 74\%. We studied the temporal feature of the backflash photon by varying the SPAD gate width and analysed the time difference in the electrical signal generated by the SPAD and SNSPD. The START signal was generated from the SPAD, and the STOP was generated by the SNSPD. Their difference is plotted in \mbox{Fig.~\ref{fig:timing-correlation}}. As varying gate width leads to a change in absolute counts for each histogram bin, the plot has been normalised with respect to peak counts. It can be seen that the spread of the curve changes significantly when the gate width is increased from 2~ns to 4~ns, beyond which the changes become minimal. 
This indicates that smaller gate widths, even in slow-gate detectors, reduce the number of backflash photons. This reduction is attributed to a shorter avalanche build-up time and the faster termination of the SPAD gate signal. For gate widths larger than 5 ns, the spread remains largely unchanged.  This suggests the maximum observed time difference between the electrical signal and the subsequent emission of backflash photons is less than 5~ns. As the SPAD is a commercial unit with front-end electronics, it doesn't allow for direct measurement of avalanche current. However, we know that the rising edge of the SPAD gate pulse shows more detections when compared with the falling edge, as it permits the carrier build-up for a longer duration before it is quenched by bringing the applied reverse bias below the breakdown voltage~\cite{gautam_spad_characterisation}. Our experiments show that the emission of backflash photons also follows the same pattern as their origin lies in the recombination of the aforementioned generated carriers.

\subsection{Eve's Learning}
Alice continuously generates encoded states and transmits them to Bob until he detects a sufficient number of photons, as determined by the block size specified in the KDE~\cite{foram_kde}. For illustration purposes, we have set the block length to 20,000.

After the completion of transmission, Alice and Bob move on to using their KDE to derive the secure keys. The first step involves the alignment of Bob's detected timestamps with Alice's bit encodings~\cite{foram_kde}. In our implementation, Bob randomly samples 2000 timestamps and the corresponding logical bit values and shares them with Alice for autocorrelations. It must be noted that the timestamps and the logical bit value are transferred over the classical authenticated channel, which is accessible to Eve.

\begin{figure}[h]
  \centering
  \begin{tabular}{@{}c@{}}
    \includegraphics[width=\linewidth]{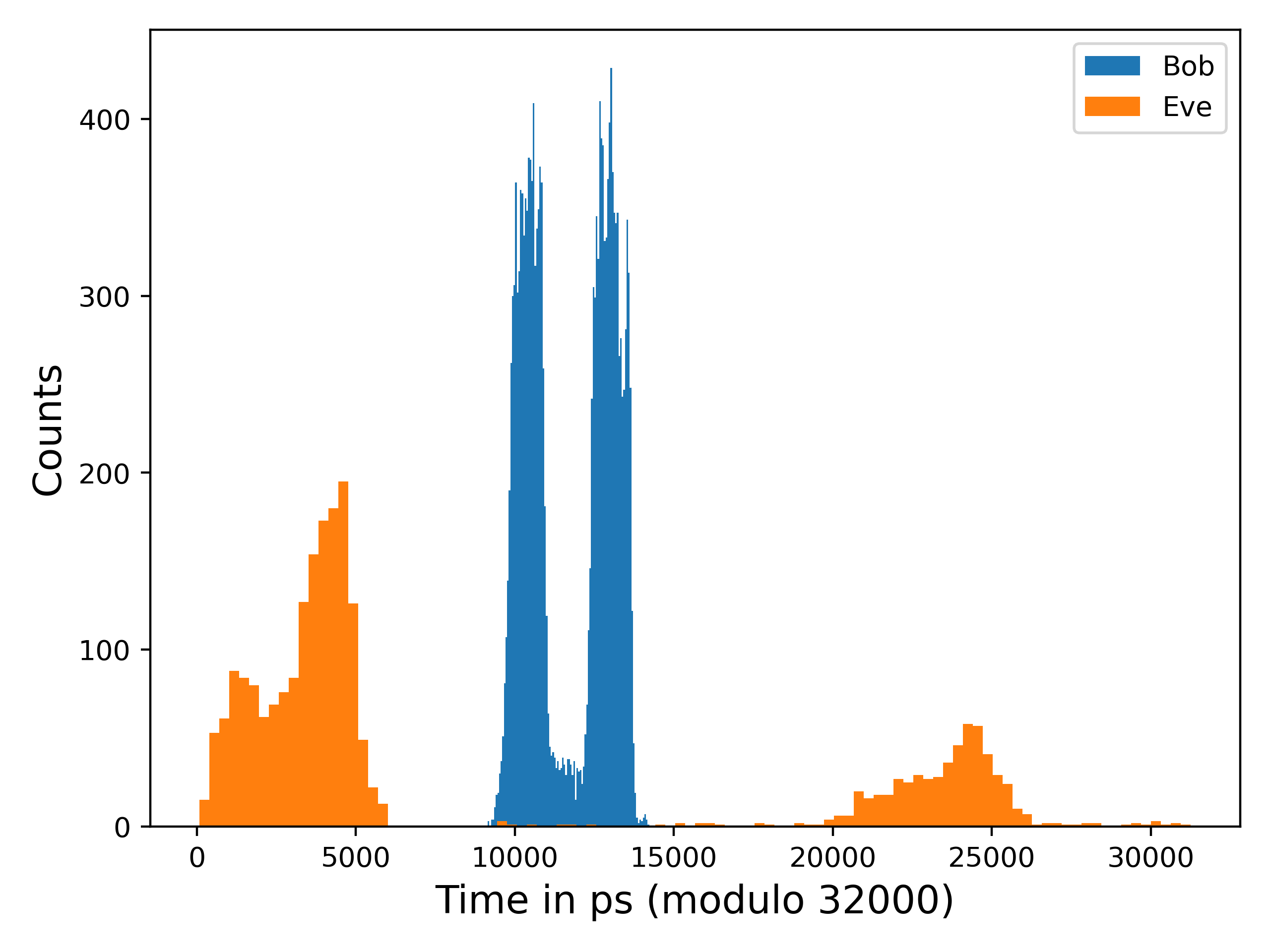} \\[\abovecaptionskip]
    \small (a) Histogram of raw timestamps
  \end{tabular}

  \vspace{\floatsep}

  \begin{tabular}{@{}c@{}}
    \includegraphics[width=\linewidth]{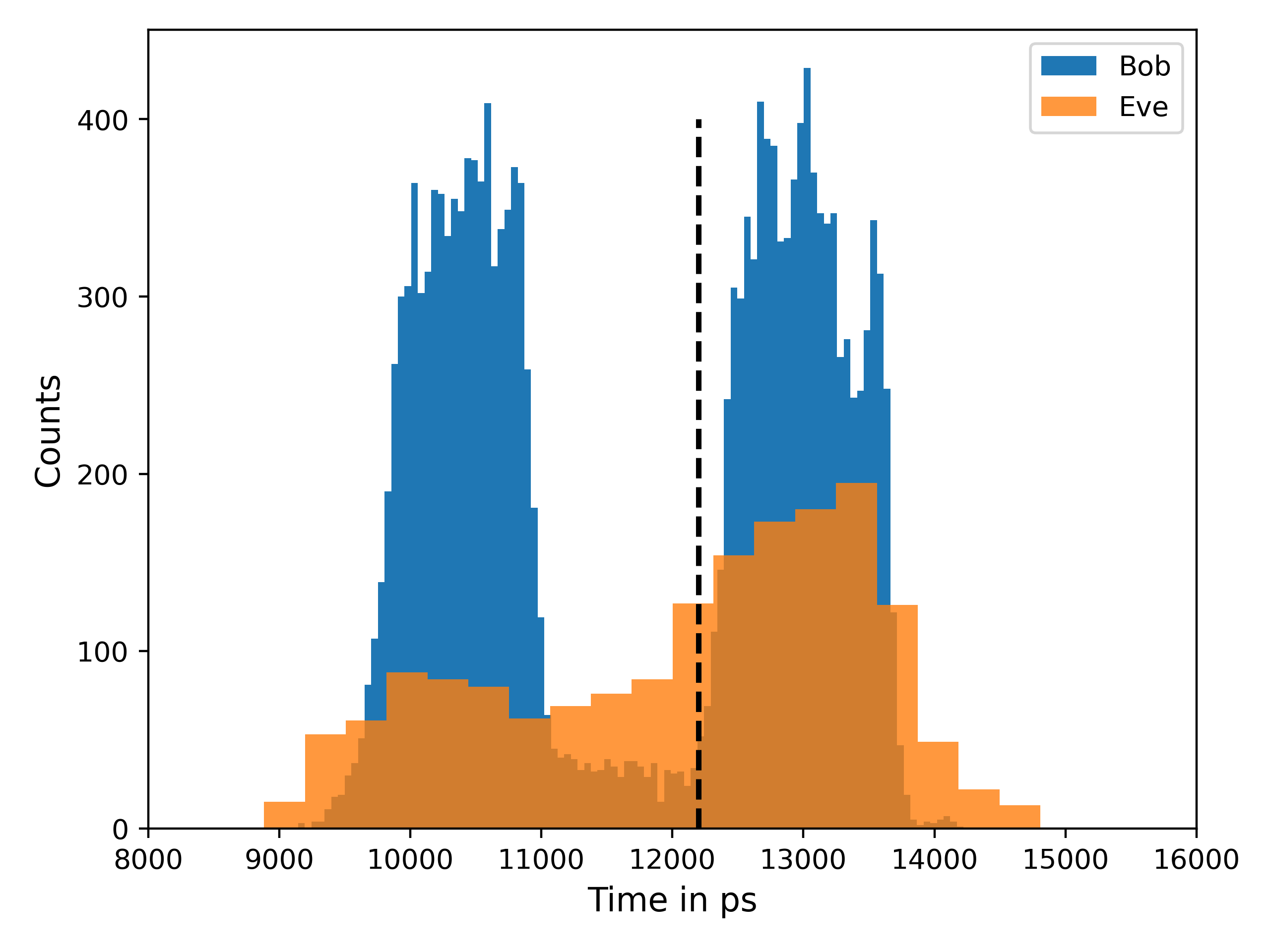} \\[\abovecaptionskip]
    \small (b) Histogram after offset calculation.
  \end{tabular}

  \caption{Histogram of Bob's and Eve's detection after a modulo with pulse repetition rate (32000~ps).}\label{fig:myfig}
  \label{fig: attack histogram}
\end{figure}

Histogram of the raw timestamp for detections at Bob and Eve are shown in \mbox{Fig.~\ref{fig: attack histogram}-(a)}. Using the timestamps and logical bit values shared by Bob for autocorrelation, Eve realises that the first cluster in her detection corresponds to the backflash photons generated by avalanche at Bob, and the second cluster corresponds to the reflections of incident pulse from the SPAD. Using the information available on the classical channel, she performs her own autocorrelation to calculate an offset, which she then applies to her timestamps, as $\text{T}_{\text{Calib}}[\text{n}]=\text{T}_{\text{raw}}[\text{n}] + \text{b}$, where $\text{T}[\text{n}]$ represents the series of timestamps detected at Eve. The low dark count rate (DCR) and high detection efficiency of the SNSPD allow her to calculate these parameters reliably. \mbox{Fig.~\ref{fig: attack histogram}-(b)} shows the histogram for Bob’s remaining 18,000 detections after excluding the bits shared for autocorrelation, along with Eve’s 1,598 detections. The overlap between Bob’s and Eve’s histogram after the shifting operation demonstrates Eve’s ability to infer Bob’s detected logical bits. The histogram at Bob revels a temporal spread of $\ket{\alpha}$ with $\ket{0}$, as shown in \mbox{Fig.~\ref{fig: attack histogram}}. In our setup, this spread can be attributed to non-return-to-zero (NRZ) encoding at Alice. The asymmetry in Eve's detection is due to delayed occurrence of backflash, as evident in \mbox{Fig.~\ref{fig:timing-correlation}}. Eve accounts for all the counts between $\ket{\alpha}_{0}$ and $\ket{\alpha}_{1}$ as belonging to $0_\text{L}$. The total counts assigned to $0_\text{L}$ and $1_\text{L}$ become 
785 and 813, respectively.

\begin{figure}
    \centering
    \includegraphics[width=\linewidth]{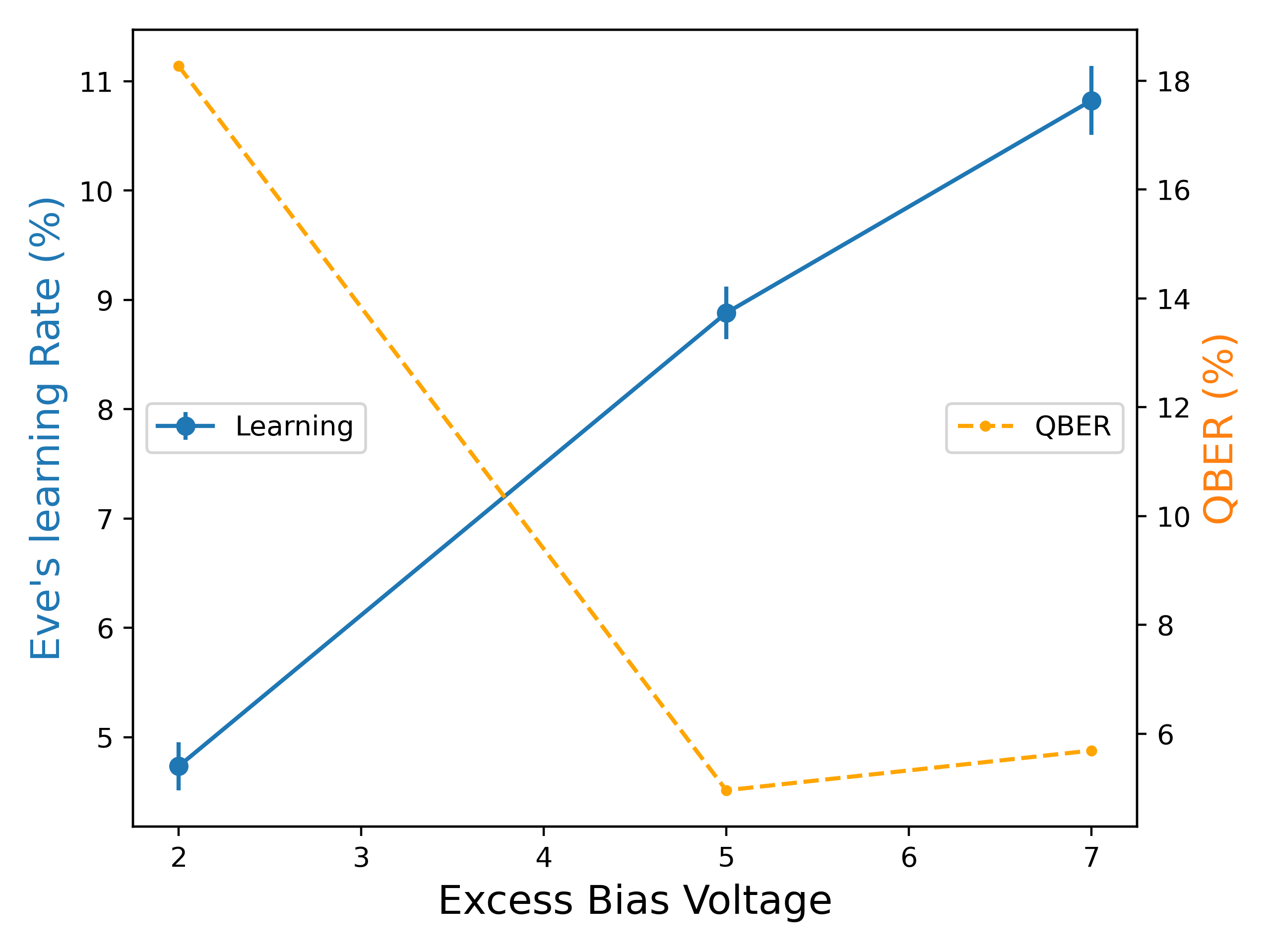}
    \caption{Effect of SPAD excess bias voltage on Eve's learning rate and QBER for QKD}
    \label{fig:learning-rate}
\end{figure}

There's an inherent trade-off between Eve's learning rate, SPAD's excess bias voltage, and QBER between Alice and Bob. Here, we would like to highlight that Eve's learning has been quantified as a ratio of Eve's detections leading to the correct identification of Bob's received bit. The result for three excess bias voltages is shown in Fig.~\ref{fig:learning-rate}. The QBER is calculated according to the sifted bits between the legitimate parties, Alice and Bob. We see that at the lowest excess bias of 2~V, where timing jitter is maximum, Eve's effective learning is minimum. Eve's learning increases with an increase in excess bias. However, our experiments showed that an increase in excess bias to 7~V led to a larger DCR at Bob's SPAD, slightly increasing QBER while also increasing Eve's learning rate. 

Our experimental setup demonstrates Eve's learning capabilities for COW-QKD protocol. It can be easily extended to other protocols that decode information depending on the time-of-flight of photons or employ passive optical components for decoding. For example, in protocols that utilise passive basis selection using beam-splitters and polarisation decoders, the backflash photons could carry the basis information for Eve. In DPS, the arrival time of photons along with the signature temporal profile of a detector could leak information~\cite{Meda2017}.

\subsection{Estimation of sifting rate and learning rate}
In this subsection, we present the theoretical estimation of the sifted key rate of the COW-QKD protocol and analyse Eve's learning rate under the backflash attack. The theoretical results are compared with the experimental findings.
As discussed in Section~\ref{section:experimental-setup}, COW-QKD employs time-bin encoding given by $0_\text{L}=\ket{\alpha}_1\otimes\ket{0}_2$ and $1_\text{L}=\ket{0}_1\otimes\ket{\alpha}_2$. Therefore, the detection of the pulse (click from SPAD) at Bob in time-bin 1 or 2 is used to sift a logic 0 or 1, respectively. Detector imperfections such as dark counts and after-pulsing probability are responsible for the error introduced. The sifting rate is then given by, 
\begin{align}
    P_{\text{sift}}=1-e^{-\lvert\alpha\rvert^2 \eta}(1-P_{\text{dark}})
\end{align}
where, $P_{\text{dark}}$ is the DCR, and $\eta$ is the channel transmittance at a given channel length. We have included the detection efficiency of SPAD into it. In an experimental setup, we strike a balance between the after-pulsing probability and hold-off time $T_{\text{hold}}$. This further reduces the sifted key rate to \cite{hadfield2009single},
\begin{align}
    P_{\text{sift}}=\frac{1-e^{-\lvert\alpha\rvert^2 \eta}(1-P_{\text{dark}})}{1+N_0\times (1-e^{-\lvert\alpha\rvert^2 \eta}(1-P_{\text{dark}}))\times T_{\text{hold}}}
\end{align}
Here, $N_0$ is the transmission rate from Alice. The QBER in the COW-QKD is defined as the probability of decoding logic 1 given logic 0 is sent or vice versa for the sifted bits. The corresponding expression is given by,
\begin{align}
    P_{\text{err}}=e^{-\lvert\alpha\rvert^2 \eta}\times (1-P_{\text{dark}})\times P_{\text{dark}}/P_{\text{sift}}
\end{align}
The probability of backflash $P_b$ can be modeled as~\cite{Meda2017},
\begin{align}
    P_b=\frac{n_{\text{Eve}}}{n_{\text{sift}}}
\end{align}
Here, $n_{\text{Eve}}$ is the number of backflash photons detected by Eve and $n_{\text{sift}}$ is the number of detections by Bob. Note that the expression of the probability of backlash is given at a photon detection efficiency of 74\% for the SNSPD. The actual backflash emission could be larger; however, Eve's learning is limited by her detector's efficiency. Using the expression of backflash probability, we can define the learning rate of Eve as,
\begin{align}
    P_{\text{learn}}^{\text{Eve}}=P_b\times P_{\text{sift}}
\end{align}
The secure key rate of COW-QKD under the backflash attack can be derived by shrinking the sifted key rate by a fraction, including the learning rate of Eve and the amount of error reconciliation needed. We define the secure key rate in this case as,
\begin{align}
    P_{\text{sec}}=P_{\text{sift}}\times (1-P_b-f\times H(e))
\end{align}
where, $H(e)$ represents the Shannon entropy of QBER, $e$. This accounts for the leakage of information in error reconciliation. $f$ is the inefficiency of the error reconciliation protocol used. The usual value for $f$ is 1.15~\cite{lin2018simple}.\par
\begin{figure}[h]
    \centering
    \includegraphics[width=\linewidth]{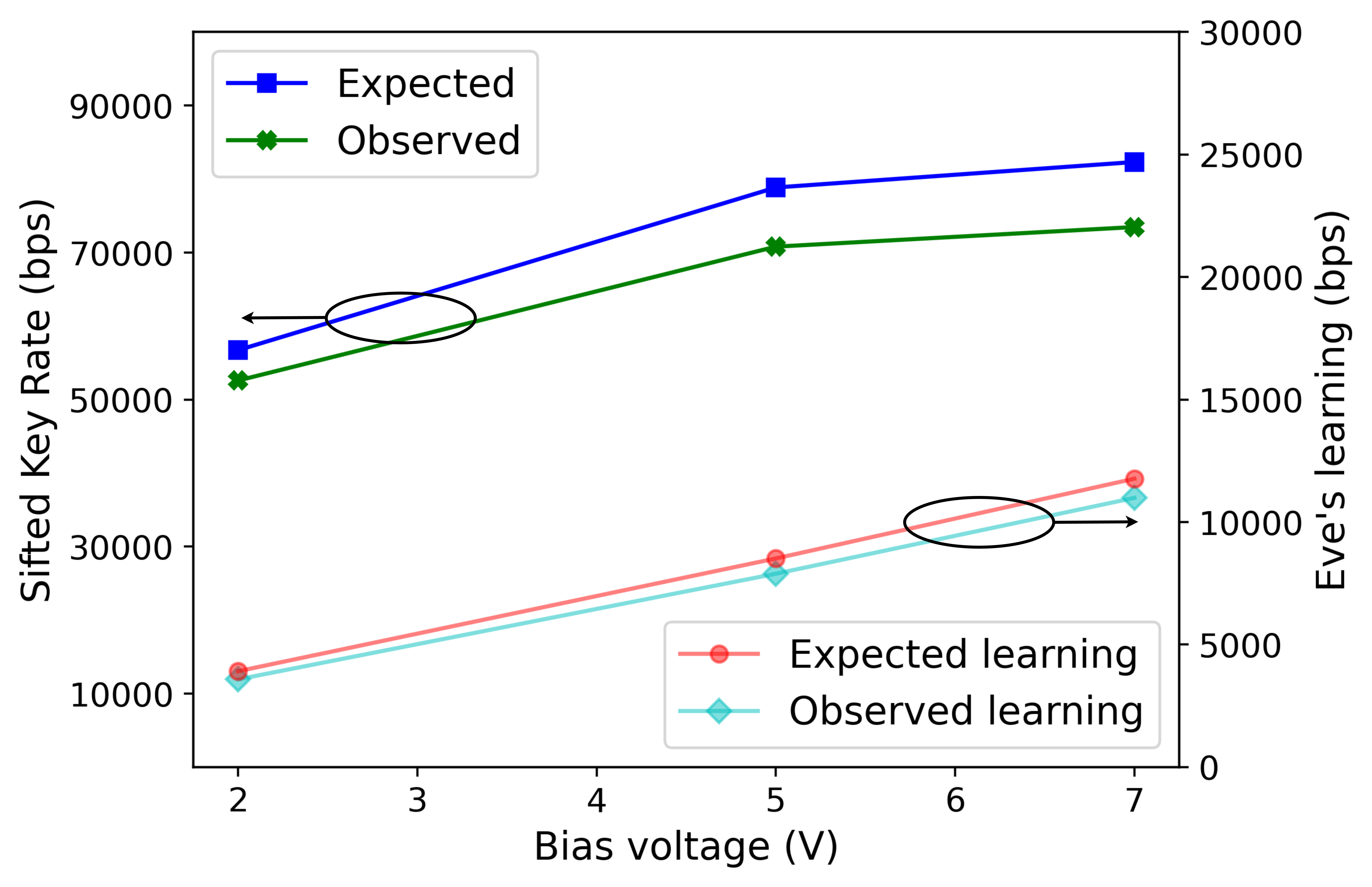}
    \caption{Comparison of expected and observed sifted key rate and learning rates at different bias voltages to SPAD.}
    \label{skr}
\end{figure}
Fig.~\ref{skr} compares the theoretical model against the experimental observations. The expected and observed sifted key rate and learning rate are compared for different excess bias voltages of SPAD. We can see that the observed sifted key rate remains below the expected sifted key rate. This can be attributed to the slight inaccuracies in ascertaining experimental parameters such as attenuation, detector efficiency, etc. As the SPAD used in our setup is a commercial unit with limited information about the inner structure of the detector, we rely on the empirically observed values of backflash probability during our experimental runs. Consequently, the expected and observed counts for Eve's learning are almost similar.

\begin{figure}[h]
    \centering
    \includegraphics[width=\linewidth]{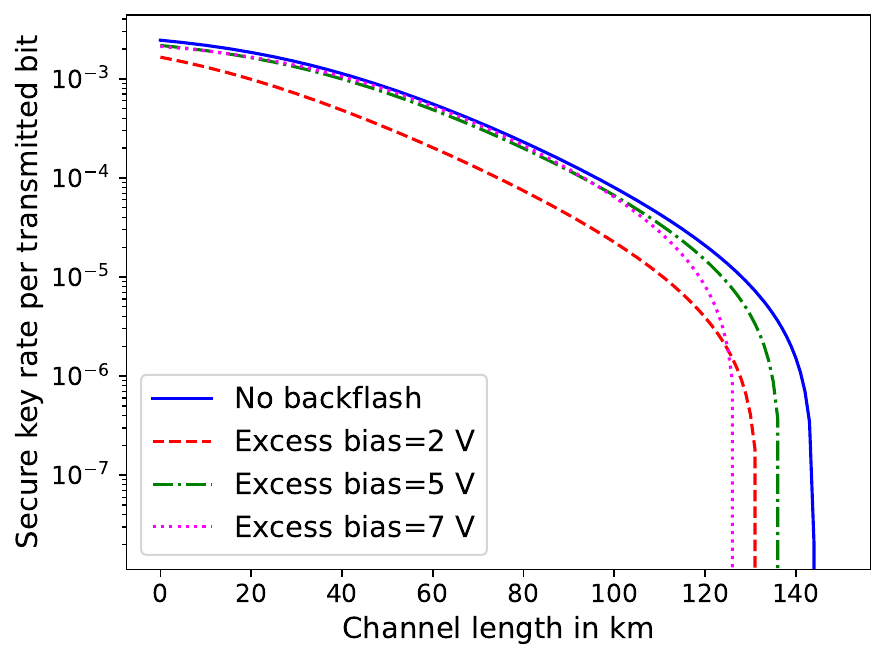}
    \caption{Comparison of secure key rate of COW-QKD with and without backflash attack. The secure key rates are estimated at various settings of bias voltages of SPAD, which affect the detection efficiency and dark count rates of SPAD.}
    \label{secukr}
\end{figure}
 
\begin{table}[h]
\caption{Parameters used for simulation.}
   \label{param}
    \renewcommand{\arraystretch}{1.5}
    \centering
    \begin{tabular}{|c|c|c|c|}
    \hline
        \textbf{Bias voltage  (in V)} & \textbf{2} & \textbf{5} & \textbf{7}\\
         \hline
         Detection efficiency (in \%) & 7 & 20 & 25 \\
         \hline
         Dark count rate (per sec.) & 100 & 200 & 350\\ 
         \hline
         Hold-off time (in $\mu$ sec.) & 10 & 10 & 10\\
         \hline
   \end{tabular}
 
\end{table}
In Fig.~\ref{secukr}, we plot a comparison of the secure key rate for COW-QKD protocol against distance. The performance is evaluated in two scenarios: with and without backflash attack. In the scenario without the backflash attack, the optimal excess bias of 5~V is chosen, yielding the highest secure key rate. As the photon detection efficiency and DCR of the SPAD varies with change in excess bias, we have analysed the effect on the secure key rate under backflash attack for three different excess bias voltages. 

The parameters used in the simulations are provided in Table~\ref{param}, with the values from the experimental setup. Based on the simulation results, the secure key rate under the backflash attack is reduced by approximately 10\% compared to the case without the attack at the same excess bias voltage of 5~V. Furthermore, changes in the excess bias voltage affect Eve's learning capabilities, which in turn impact the secure key rate. 

From a practical standpoint, we studied the effect of finiteness on the final secure key. In its nature, the attack is passive as Eve does not interact with the quantum states, and introduces no additional error. The finite size effects, therefore, includes statistical fluctuation in estimated QBER and failure probability of error correction and privacy amplification, which can be estimated as,
\begin{equation}
    P_{\text{sec}}^{\text{finite}}=P_{\text{sift}}\times (1-\text{leak}_{\text{EC}}-\beta_{\text{EC}}-\beta_{\text{PA}}).
\end{equation}
Here, $\text{leak}_{\text{EC}}$ represents the information leaked during error correction, and $\beta_{\text{EC}}$ the probability of failure of error correction, and $\beta_{\text{PA}}$ is the information gained by Eve and probability of failure of privacy amplification \mbox{\cite{walenta2014fast}}.
\section{Conclusion}
In this work, we have successfully demonstrated the vulnerability of a COW-QKD setup to backflash attacks by experimentally showcasing Eve's ability to exploit backflash photons emitted by SPADs. Our results show that Eve can gain significant information about the sifted key on a QKD link by analyzing these backflash photons. Additionally, we studied the impact of SPAD’s excess bias voltage on both QKD performance and Eve’s learning rate. The findings indicate that Eve’s ability to infer Bob’s detection events increases with increasing excess bias voltage.

Mitigating the backflash attack is essential for the secure implementation of QKD protocols. Experimentally, this vulnerability can be addressed by placing optical isolators or spectral filters before the SPAD or by replacing SPADs with superconducting nanowire single-photon detectors at Bob’s end. At the protocol level, measurement-device-independent (MDI) QKD and device-independent (DI) QKD provide robust solutions, inherently rendering backflash photons inconsequential.

In summary, achieving the information-theoretic security promised by QKD requires addressing vulnerabilities arising from device imperfections. This work highlights the importance of accounting for these challenges in experimental setups and protocol design to ensure the robustness of QKD systems against practical attacks.
\section*{Acknowledgments}
This work was partly supported by the Mphasis F1 foundation. AKS and VPS would like to thank Mohd. Razin Ashfaque for the discussion regarding the FPGA setup. AKS and NL would like to thank Aryan Bhardwaj, Anagha Gayathri, and Valliamai Ramanathan for their inputs on the manuscript.
\bibliographystyle{ieeetr}
\bibliography{sample}

\end{document}